\documentclass[aps,preprint,prl,showpacs]{revtex4}
\usepackage{graphicx}
\usepackage{subfigure}
\usepackage{braket}
\usepackage[T1]{fontenc} 
\usepackage{epsfig,latexsym}
\usepackage{color}
\usepackage[latin1]{inputenc}
\usepackage{amsmath}
\usepackage{amssymb}
\usepackage{amsfonts}
\usepackage{indentfirst}
\usepackage{hyperref}

\def\d{\mathrm{d}}

\begin{document}
\title{Indirect evidence for the Gouy phase for matter waves}
\author{I. G. da Paz\footnote{Corresponding author.}}
\affiliation{Departamento de Matemática, Universidade Federal do
Piauí, Campus Senador Helvídio Nunes de Barros, Picos, PI 64600-000,
Brazil}\affiliation{Departamento de F\'{\i}sica, Instituto de
Ci\^encias Exatas, Universidade Federal de Minas Gerais, Caixa
Postal 702, Belo Horizonte, MG 30123-970, Brazil, e-mail:
irismar@fisica.ufmg.br, Phone: +55-31-3499-5605, Fax:
+55-31-3499-5688.}
\author{M. C. Nemes}
\affiliation{Departamento de F\'{\i}sica, Instituto de Ci\^encias
Exatas, Universidade Federal de Minas Gerais, Caixa Postal 702, Belo
Horizonte, MG 30123-970, Brazil}
\author{C. H. Monken}
\affiliation{Departamento de F\'{\i}sica, Instituto de Ci\^encias
Exatas, Universidade Federal de Minas Gerais, Caixa Postal 702, Belo
Horizonte, MG 30123-970, Brazil} \affiliation{Huygens Laboratory, P.
O. Box 9504, 2300 RA Leiden, The Netherlands}
\author{S. P\'adua}
\affiliation{Departamento de F\'{\i}sica, Instituto de Ci\^encias
Exatas, Universidade Federal de Minas Gerais, Caixa Postal 702, Belo
Horizonte, MG 30123-970, Brazil}
\author{J. G. Peixoto de Faria}
\affiliation{Departamento de F\'{\i}sica e Matem\'atica, Centro
Federal de Educa\c{c}\~ao Tecnol\'ogica de Minas Gerais, Av.
Amazonas 7675, Belo Horizonte, MG 30510-000, Brazil}

\begin{abstract}
We show that the well known geometric phase, the Gouy phase in
optics can be defined for matter waves in vacuum as well.  In
particular we show that the underlying physics for the ``matter
waves'' Gouy phase is the generalized Schrödinger-Robertson
uncertainty principle, more specifically, the off diagonal elements
of the covariance matrix. Recent experiments involving the
diffraction of fullerene molecules and the uncertainty principle are
shown to be quantitatively consistent with the existence of a Gouy
phase for matter waves.
\end{abstract}


{\it Accepted for publication in PLA}

\maketitle The extra phase shift experienced by a converging light
wave passing trough its focal point, also known as the Gouy phase
shift, is a well known phenomenon in optics.  Since its observation
reported by Gouy in 1890 \cite{gouy1,gouy2}, this phase shift
(sometimes referred to as a phase anomaly), its physical origin and
its consequences have been objects of study, especially in the last
decades \cite{boyd,simon,hariharan,feng98,feng01,yang}. Although it
is often presented as a property of Gaussian light beams
\cite{siegman} the Gouy phase shift appears in any kind of wave that
is submitted to some sort of transverse spatial confinement, be it
by focusing or by diffraction through small apertures. As discussed
in \cite{feng01}, when a wave is focused, the Gouy phase shift is
associated to the propagation from $-\infty$ to $+\infty$ and is
equal to $\pi/2$ for cylindrical waves (line focus), and $\pi$ for
spherical waves (point focus).  In the case of diffracted waves, the
Gouy phase shift is associated to the propagation from the
diffraction aperture to infinity, and amounts to $\pi/4$ ($\pi/2$)
for one-dimensional (two-dimensional) apertures.

The Gouy phase shift has been observed in water waves
\cite{chauvat}, acoustic \cite{holme}, surface plasmon-polariton
\cite{zhu}, and phonon-polariton \cite{feurer} pulses.  In this work
we show that it can also be observed in matter waves, despite the
abstract nature of the latter, and is directly related to the
generalized uncertainty principle. Some effects related to the Gouy
phase shift in electromagnetic waves, such as pulse reshaping
\cite{horvat} and acceleration \cite{horvath}, anomalous spectral
behavior \cite{gbur}, among others, rise interesting questions about
their possible counterparts in matter waves.

We start our analysis by taking the simple route of a direct
comparison between the Gaussian solutions of the paraxial wave
equation and the two-dimensional Schr\"odinger equation. Later, we
focus on the one-dimensional problem of diffraction of massive
particles through a single slit. We take as an example a recent
diffraction experiment with fullerene molecules
\cite{nairz02,nairz03} and show that the Gouy phase shift can be
inferred from the reported experimental data.

Consider a stationary electric field in vacuum
\begin{equation}
E(\vec{r})=A(\vec{r})\exp(ikz).
\label{campoeletrico}
\end{equation}
The paraxial approximation consists in assuming that the complex
envelope function $A(\vec{r})$ varies slowly within one wavelength
$\lambda_L=2\pi/k$. Under this condition the equation for
$A(\vec{r})$ can be immediately obtained and reads \cite{Saleh}
\begin{equation}
\left(\frac{\partial^{2}}{\partial
x^{2}}+\frac{\partial^{2}}{\partial y^{2}}+i4\pi
\frac{1}{\lambda_{L}}\frac{\partial}{\partial
z}\right)A\left(x,y,z\right)=0. \label{paraxial}
\end{equation}

Consider now the two-dimensional Schrödinger equation for a free
particle of mass $m$
\begin{equation}
\left(\frac{\partial^{2}}{\partial
x^{2}}+\frac{\partial^{2}}{\partial
y^{2}}+2i\frac{m}{\hbar}\frac{\partial}{\partial
t}\right)\psi(x,y,t)=0. \label{schrodinger}
\end{equation}
Here, $\psi(x,y,t)$ stands for the wave function of the particle in
time $t$. Assuming that the longitudinal momentum component $p_z$ is
well-defined \cite{Viale}, i.e., $\Delta p_z \ll p_z$, we can
consider particle's movement in the $z$ direction is classical and
its velocity in this direction remains constant. In this case one
can interpret the time variation as the variation in this direction
according to the relation $t=z/v_{z}$. Now using the fact that
$\lambda_{P}=h/p_z$ and substituting in Eq. \eqref{schrodinger} we
get
\begin{equation}
\left(\frac{\partial^{2}}{\partial
x^{2}}+\frac{\partial^{2}}{\partial y^{2}}+i4\pi
\frac{1}{\lambda_{P}}\frac{\partial}{\partial
z}\right)\psi\left(x,y,t=z/v_{z}\right)=0. \label{schodinger2}
\end{equation}
The condition of constant velocity in the propagation direction has
been used in the analysis of recent diffraction experiments with
fullerene molecules \cite{nairz02,nairz03}. A theoretical model
using this assumption was also shown to qualitatively reproduce the
data well \cite{Viale}.

We show the particle's counterpart using the initial gaussian state
\begin{equation}
 \psi\left(x,y,0\right)=\left(\frac{1}{b\sqrt{\pi}}\right)
 \exp\left[-\frac{(x^{2}+y^{2})}{b^{2}}\right],
 \label{initial}
\end{equation}
we get for an arbitrary time $t$ \cite{Paz}
\begin{eqnarray}
\psi\left(x,y,t\right)&=&\left[\frac{1}{B\left(t\right)\sqrt{\pi}}\right]
\exp\left(
-\frac{x^{2}+y^{2}}{B^{2}\left(t\right)}\right)\nonumber\\
&\times&\exp\left\lbrace
i\left[\frac{m\left(x^{2}+y^{2}\right)}{2\hbar
R\left(t\right)}+\mu\left( t\right)\right]\right\rbrace,
\label{pacotegaussiano}
\end{eqnarray}
where $\mu(t)$ is the analogous to the Gouy phase. The comparison
with the solution of the wave equation in the paraxial approximation
with the same condition at $z=0$ yields
\begin{equation}
    B\left(t\right)=
    b\left[1+\left(\frac{t}{\tau_{b}}\right)^{2}\right]^{\frac{1}{2}},
    \label{largurafeixeparticula}
\end{equation}
\begin{equation}
    R\left(t\right)=t\left[1+\left(\frac{\tau_{b}}{t}\right)^{2}\right],
    \label{raiofeixeparticula}
\end{equation}
\begin{equation}
    \mu\left(t\right)=-\arctan\left(\frac{t}{\tau_{b}}\right)
    \label{fasegouyparticula},
\end{equation}
and
\begin{equation}
\tau_{b}=\frac{mb^{2}}{\hbar}.
\end{equation}
Notice that the parameter $\tau_{b}$ is related to the initial
condition only and is responsible for two regimes of the growing
width $B(t)$, in close analogy to Rayleigh range which separates the
width $w(z)$ in two qualitatively different regimes, as is well
known \cite{Saleh}.

Next we show that $\mu(t)$ is directly related to the
Schrödinger-Robertson generalized uncertainty principle. For
quadratic evolutions (as the free evolution in the present case) the
determinant of the covariance matrix saturates to its minimum value,
\begin{equation}
\det\left(
\begin{array}{cc}
{\sigma_{xx}} & {\sigma_{xp}} \\
{\sigma_{xp}} & {\sigma_{pp}}
\end{array}
\right)=\frac{\hbar^{2}}{4}
\end{equation}
where $\sigma_{xx}=\langle
\hat{x}^{2}\rangle-\langle\hat{x}\rangle^{2}=\frac{B(t)^{2}}{2}$,
$\sigma_{pp}=\langle \hat{p}^{2}\rangle-\langle
\hat{p}\rangle^{2}=\frac{\hbar^{2}}{2b^{2}}$, and
$\sigma_{xp}=\frac{1}{2}\langle
\hat{x}\hat{p}+\hat{p}\hat{x}\rangle-\langle \hat{x}\rangle \langle
\hat{p}\rangle=\frac{\hbar t}{2\tau_{b}}$. Since the covariance
$\sigma_{xp}$ is non-null if the gaussian state exhibits squeezing
\cite{souza2008}, if one measures $\sigma_{xp}$, from the above
relation it is possible to infer the Gouy phase for a matter wave
which can be described by an evolving coherent wave packet.

Now an important question is in order: What does experiment say?

In Ref. \cite{nairz02} an experimental investigation of the
uncertainty principle in the diffraction of fullerene molecules is
presented. In that experiment, a collimated molecular beam crosses a
variable aperture slit and its width is measured as a function of
the slit width (see Fig. 3 in Ref. \cite{nairz02}).

As discussed in Ref. \cite{Viale}, given the way the fullerene
molecules are produced, it is reasonable to assume that the outgoing
beam after the diffraction slit has a random transverse momentum.
Here, we suppose that the transverse momentum follows a gaussian
distribution with zero mean and width $\delta k_x$, that is related
to the experimental angular divergence of the beam, as explained in
Ref. \cite{Zeilinger3}. So, the state of fullerene molecules which
leave the slit with width $b$ is given by (in transverse direction)
\begin{equation}
\rho(x,x^{\prime},0)=\frac{1}{b\sqrt{\pi}}\int
\exp\left(-\frac{x^{2}+x^{\prime 2}}{2b^{2}}\right)
\exp\left[ik_{x}(x-x^{\prime})\right] g(k_{x})dk_{x}
\label{initial_mix}
\end{equation}
where
\begin{equation}
g(k_{x})=\frac{1}{\sqrt{\pi}\delta k_{x}}
\exp\left(-\frac{k_{x}^{2}}{\delta k_{x}^{2}}\right),
\end{equation}
is a gaussian probability distribution function for the transverse
momentum with width $\delta k_{x}/\sqrt{2}$.

At $t=\frac{(z=L)}{v_{z}}$, where $v_{z}$ is the most probable
velocity in $z$ direction, the calculation of the time evolution is
straightforward and the elements of the covariance matrix are now
given by
\begin{equation}
\sigma_{xx}
=\frac{B(t)^{2}}{2}\left[1+\left(\frac{\tau_{b}B(t)\delta
k_{x}}{R(t)}\right)^{2}\right]
\equiv \frac{\bar{B}\left(t\right)^2}{2}, \label{sigmaxx}
\end{equation}
\begin{equation}
\sigma_{pp} =\frac{\hbar^{2}}{2}\frac{(1+b^{2}\delta
k_{x}^{2})}{b^{2}},
\end{equation}
\begin{equation}
\sigma_{xp}=\frac{\hbar}{2}\left(\frac{t}{\tau_{b}}\right)\left(1+b^2\delta
k_{x}^{2}\right).
\end{equation}
In Eq. \eqref{sigmaxx}, we define $\bar{B}\left(t\right)$ as the
counterpart of the width $B\left(t\right)$ for the partially
coherent gaussian state given in Eq. \eqref{initial_mix}. The
effective detected intensity is a convolution of the detector
resolution function $D(x)$ and the real intensity $I(x,t)$ (where
$I(x,t)=\rho(x,x,t)$)\cite{nairz02,Viale}.

For pure states, the Gouy phase $\mu\left(t\right)$ and the width
$B\left(t\right) $ are related by the expression \cite{feng01}
\begin{equation}
    \mu\left(t\right) = -\frac{\hbar}{2m}\int^t \frac{\d t}{B\left(t\right)^2}.
    \label{mu_et_B}
\end{equation}
We conjecture that this relation holds for partially coherent states. So, for the state
given in Eq. \eqref{initial_mix}, the Gouy phase is
\begin{equation}
    \mu(t)
    =-\frac{1}{2\sqrt{1+b^2\delta k_x^2}}
    \arctan\left(\frac{2\sigma_{xp}}{\hbar\sqrt{1+b^{2}\delta
    k_{x}^{2}}}\right).
    \label{fasedacorrelacao}
\end{equation}
Note that, again $\mu(t)$ is related to $\sigma_{xp}$ and is
affected by the partial coherence of the initial wave packet.

Furthermore, as discussed before from the saturation of the
determinant of the covariance matrix we get for $\sigma_{xp}$,
\begin{equation}
\sigma_{xp}=\frac{\hbar}{2}\sqrt{1+b^{2}\delta
k_{x}^{2}}\left[\left(\frac{W_{FWHM}}{2\sqrt{\ln2}b}\right)^{2}-1\right]^{\frac{1}{2}},
\end{equation}
where $W_{FWHM}$ is the full width at half maximum of the
diffraction pattern at the screen. The experimental results for
$W_{FWHM}$ as reported in Ref. \cite{nairz02} is shown in Fig. 1 and
compared with our theoretical calculation Eq. \eqref{sigmaxx} (where
$W_{FWHM}=2\sqrt{2\ln2\sigma_{xx}}$). We use $b\rightarrow b/3$ due
to van der Waals forces for the smallest slit width
\cite{nairz02,Toennies,Viale} and $D=12\;\mathrm{\mu m}$ for the
spatial resolution of the detector \cite{nairz02}. The curve for
$\sigma_{xp}$ is shown in Fig. 2
\begin{figure}[htp]
    \centering
    \includegraphics[width=7 cm]{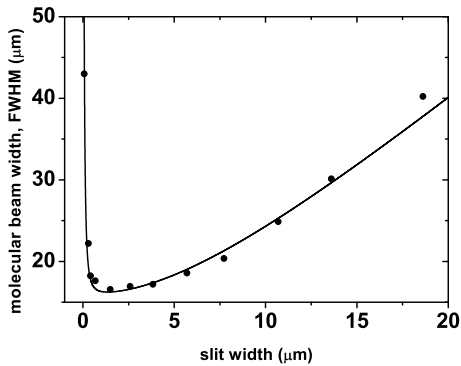}\\
    \caption{Width of a beam of fullerene $\mathrm{C}_{70}$ molecules
as a function of the slit width. The solid curve corresponds to our
calculation and the points are the results obtained in the
experiment reported in Ref. \cite{nairz02}. We used $\delta
k_{x}=9.0\times10^{6}\:\mathrm{m^{-1}}$ and $t=6.65\:\mathrm{ms}$.}
\end{figure}

\begin{figure}[htp]
    \centering
    \includegraphics[width=7 cm]{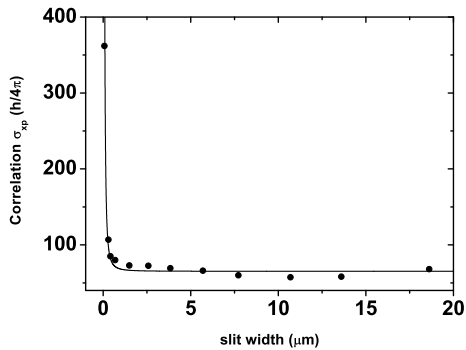}\\
    \caption{$x-p$ correlation as a function of the slit width. The
    solid curve corresponds to our calculation and the points were obtained from the experimental results
    reported in Ref. \cite{nairz02}.  The parameters are the same as in
    Fig. 1.}
\end{figure}
Now we show that this experiment is consistent with the existence of
a phase. In Fig. 3 we show the phase, taken from Eq.
\eqref{fasedacorrelacao}. As expected the phase variation is $\pi/4$
\cite{feng01}.

The parameter which measures the partial coherence in transverse
direction of the beam is given by $\delta k_{x}$. By fitting the
data in Fig. 1 we obtained $\delta
k_{x}=9.0\times10^{6}\:\mathrm{m^{-1}}$. It corresponds to an
angular aperture of $2.4\:\mathrm{\mu rad}$. This is completely
compatible with the experimental value quoted in Ref.
\cite{Zeilinger3} ($2\leq\theta\leq10\:\mathrm{\mu rad}$).

\begin{figure}[htp]
    \centering
    \includegraphics[width=7 cm]{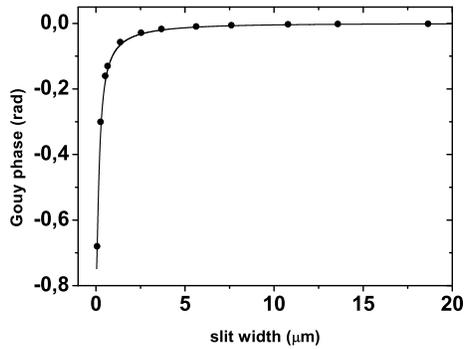}\\
    \caption{Gouy phase as a function of the slit width. The solid curve
    corresponds to our calculation and the points were obtained from the experimental results
    reported in Ref. \cite{nairz02}.  The parameters are the same as in
    Fig. 1.}
\end{figure}

In summary we have shown that a geometrical phase for matter waves
can be defined and shown to be compatible with existing experimental
data. We have also given interpretation as to the physical content
of this phase for matter waves: it is intimately related to the
uncertainty principle, more specifically to the off diagonal element
of the covariance matrix, $\sigma_{xp}$. We have also shown that the
existence of this phase is compatible with experimental data
involving the diffraction of fullerene molecules. We hope our
results encourage experimentalists to implement a direct measure of
this phase, since nowadays the fabrication of lenses capable of
focusing matter beams are available \cite{Paz1}. Of course the
experiment with fullerene molecules is not the best candidate for
exhibiting the phase due to its incoherence in transverse direction.
In this aspect experiment with neutrons should be ``cleaner",
unfortunately not available yet.
\begin{acknowledgments}
We would like to thank K. M. Fonseca Romero for a careful reading of
the manuscript. This work was partially supported by CNPq.
\end{acknowledgments}

\end{document}